\documentclass[prc,preprint,tightenlines,showpacs]{revtex4}
\usepackage{amsmath}
\usepackage{graphicx}
\usepackage{graphics}
\begin{document}
\title{Low-energy operators in effective theories}
\author{C. Felline$^{1}$, N.P. Mehta$^{2}$, 
        J. Piekarewicz$^{1}$, and J.R. Shepard$^{2}$}
\affiliation{${}^{1}$Department of Physics, Florida State University, 
             Tallahassee, FL 32306 \\
	     ${}^{2}$Department of Physics, 
             University of Colorado, Boulder, CO 80309}
\date{\today}

\begin{abstract}
Modern effective-theory techniques are applied to the nuclear
many-body problem. A novel approach is proposed for the
renormalization of operators in a manner consistent with the
construction of the effective potential.  To test this approach, 
a one-dimensional, yet realistic, nucleon-nucleon potential is
introduced. An effective potential is then constructed by tuning its
parameters to reproduce the exact effective-range expansion and a
variety of bare operators are renormalized in a fashion compatible
with this construction. Predictions for the expectation values of
these effective operators in the ground state reproduce the results of
the exact theory with remarkable accuracy (at the 0.5\% level). This
represents a marked improvement over a widely practiced approach that
uses effective interactions but retains bare operators. Further,
it is shown that this improvement is more impressive as the operator
becomes more sensitive to the short-range structure of the
potential. We illustrate the main ideas of this work using the elastic
form factor of the deuteron as an example.
\end{abstract}
\pacs{}
\maketitle

\section{Introduction}
\label{introduction}
The construction of effective interactions for use in shell-model
studies of nuclear structure has enjoyed a resurgence due in part to
the development of modern effective-field
theories~\cite{Lep97,Bea00}. Further, tremendous advances in raw
computational power and numerical techniques have enabled the
consistent and systematic implementation~\cite{Nav00a,Nav00b,Fay01} of
25-year-old approaches based on the so-called
similarity-transformations methods~\cite{Kre74,Suz80,Suz82,Suz83}.
Such implementations bypass most of the recent criticism levied on
frequently employed shell-model approaches that rely on effective
interactions that do not follow in any systematic way from a realistic
nucleon-nucleon ($NN$) interactions\cite{Hax99,Hax01}. Indeed,
similarity-transformations methods indicate how bare interactions and
operators should be modified in a systematic way to account for the
inevitable effects of truncations.

Earlier work by two of us (JP and JRS) on low-dimensional quantum
magnets~\cite{Pie96,Pie98a,Pie98b,Pie98c,Pie99} made us familiar 
with a variety of theoretical approaches that have been recently
adapted to the nuclear many-body problem~\cite{Mue02}. In that work 
it was shown how to combine similarity-transformation methods,
specifically the COntractor REnormalization (CORE) approach of
Morningstar and Weinstein\cite{Mor94,Mor96}, with effective
interactions methods, such as those discussed by
Lepage\cite{Lep97}. In particular, accurate predictions for the
ground-state energy of the three-body system were made with relatively
little computational effort when both techniques were used in a
complementary fashion.  As discussed in other recent
publications~\cite{Bog01}, these similarity transformation methods may
be understood in the context of effective theories, which in turn rely
on renormalization-group techniques~\cite{Bir99}.

What it is not at all clear (at least to us) in effective-theory
approaches, is how to modify operators in a manner consistent to the
modifications of the underlying Hamiltonian. The need for consistently
modified operators must be emphasized. Parametrized operators are
often added to improve quantitative agreement with data, but their
origin is left unclear. Even in approaches based on similarity
transformations where the modification to operators is well
delineated, it remains common practice to employ bare (rather than
renormalized) operators. In this work we adopt some modern concepts of
low-energy Effective Theories (ET's) in the hope of improving some of
these shortcomings.  The basic assumption of ET's is that the
complicated, and likely unknown, short-distance details of a theory
are hidden from a long-wavelength probe. It should then be possible to
modify the corresponding portion of the potential leaving its low
energy properties intact. In order to achieve this, the low-energy
properties must be known in advance either from experiment or, as in
the case of this study, from solving the bare theory exactly at low
energy.  As has been observed by many authors~\cite{Bed02,Bar02}, ET's
for the $NN$ interaction reproduce the effective range theory of many
decades ago. Part of the inspiration for the operator methods reported
here arose from an especially simple derivation of effective-range
theory which appears in the texts by Schiff and
Taylor~\cite{Sch68,Tay72}.

The paper has been organized as follows. In Sec.~\ref{formalism}
a simple derivation of the effective-range expansion in one spatial
dimension is presented. Next, a prescription for the renormalization
of effective operators that is consistent with the construction of 
the effective potential is introduced. In Sec.~\ref{results} 
expectation values for various effective operators are computed and 
are then compared to those obtained in the bare (exact) theory.
Finally, conclusions and some ideas for the future are discussed in 
Sec.~\ref{conclusions}.

\section{Formalism}
\label{formalism}

The aim of this section is to adapt a textbook derivation of the
effective-range formula in three spatial
dimensions~\cite{Sch68,Tay72} to the one-dimensional problem 
considered here. These ideas are then used for the construction
of an effective interaction that reproduces the scattering length
and effective range of the exact ({\it i.e.,} bare) theory.
Finally, an approach is proposed for the renormalization of 
operators in a manner which is consistent with the construction 
of the effective interaction.

\subsection{Effective-range formula}

To arrive at the effective-range formula we proceed along the lines 
of Schiff and Taylor~\cite{Sch68,Tay72}, adapting their derivation 
to the one-dimensional case considered here. The even-parity solution 
of the scattering problem satisfies the time-independent Schr\"odinger
equation
\begin{equation}
 \left[\frac{d^2}{dx^2}+k^2-U(x)\right]\psi_{k}(x)=0 \quad
 \Big(k^{2}\equiv 2\mu E \hspace{0.25cm} {\rm and} 
  \hspace{0.25cm} U(x)\equiv2\mu V(x)\Big)\;, 
 \label{SchrEq}
\end{equation}
subject to the following boundary conditions:
\begin{subequations}
\begin{eqnarray}
 \lim_{x\rightarrow 0}\psi_{k}(x)
 \hspace{-0.05in}&=&\hspace{-0.04in}
  1+{\cal O}(x^{2})\;, \label{BoundConda}\\
 \lim_{x\rightarrow \infty}\psi_{k}(x)
 \hspace{-0.05in}&=&\hspace{-0.04in}
 \phi_{k}(x)\equiv
 \cos(kx)-\tan\delta(k)\sin(kx)=
 \frac{\cos\left(kx+\delta(k)\right)}{\cos{\delta(k)}}\;.
 \label{BoundCondb}
\end{eqnarray}
\label{BoundCond}
\end{subequations}
Note that $\phi_{k}(x)$ denotes the solution of the free Schr\"odinger
($U(x)\equiv 0$) equation that coincides with $\psi_{k}(x)$ at large 
$x$. It then follows immediately from Schr\"odinger's equation that
\begin{subequations}
\begin{eqnarray}
 \frac{dW(\psi_{k},\psi_{0})}{dx}
       \hspace{-0.04in}&=&\hspace{-0.04in}
       k^{2}\psi_{k}(x)\psi_{0}(x) \;, \\
 \frac{dW(\phi_{k},\phi_{0})}{dx}
       \hspace{-0.04in}&=&\hspace{-0.04in}
       k^{2}\phi_{k}(x)\phi_{0}(x) \;,
\end{eqnarray}
\label{dWronskian}
\end{subequations}
where the Wronskian of $f$ and $g$ is defined as
\begin{equation}
  W(f,g)(x)\equiv 
  \left|\begin{matrix}
     f(x)          & g(x)          \cr
     f^{\prime}(x) & g^{\prime}(x) 
 \end{matrix}\right|=
\Big[f(x)g^{\prime}(x)-f^{\prime}(x)g(x)\Big] \;.
 \label{Wronskian}
\end{equation}
Upon integrating the difference of Eqs.~(\ref{dWronskian}) 
one obtains
\begin{equation}
  \Big[W(\phi_{k},\phi_{0})(x)-
  W(\psi_{k},\psi_{0})(x)\Big]_{0}^{\infty}=
  k^{2}\int_{0}^{\infty} dx\Big(\phi_{k}(x)\phi_{0}(x)-
                           \psi_{k}(x)\psi_{0}(x)\Big)\;. 
\label{EffRange1}
\end{equation}
The contribution from the upper limit of the integral to the 
left-hand side of the equation vanishes, as 
$\psi_{k}(x)\!=\!\phi_{k}(x)$ at large distances. Further, as 
the derivative of the exact scattering solution vanishes at 
$x=0$ [see Eq.~(\ref{BoundConda})] the Wronskian 
$W(\psi_{k},\psi_{0})$ vanishes as well. This yields
\begin{equation}
  W(\phi_{k},\phi_{0})(x\!=\!0)=
  \Big[\phi_{k}(0)\phi^{\prime}_{0}(0)-
       \phi_{k}^{\prime}(0)\phi_{0}(0)\Big]=
  k^{2}\int_{0}^{\infty} dx\Big(\phi_{k}(x)\phi_{0}(x)-
                           \psi_{k}(x)\psi_{0}(x)\Big)\;,
\label{EffRange2}
\end{equation}
which in turn generates the well known effective-range formula
\begin{equation}
  k\tan\delta(k)=\frac{1}{a_{0}}-
  k^{2}\int_{0}^{\infty} dx\Big(\psi_{k}(x)\psi_{0}(x)-
                           \phi_{k}(x)\phi_{0}(x)\Big)
		=\frac{1}{a_{0}}-\frac{r_{0}}{2}k^{2}
		+{\cal O}(k^{4})\;.	   
\label{EffRange3}
\end{equation}
Note that the (even-parity) scattering length and
effective range parameters have been defined as
\begin{equation}
 a_{0}^{-1}=
 \lim_{k\rightarrow 0}k\tan\delta(k)\;, 
 \quad
 r_{0}=
 2\int_{0}^{\infty} dx\Big(\psi_{0}^{2}(x)-
                           \phi_{0}^{2}(x)\Big)\;.
\label{EffParams}
\end{equation}

\subsection{Effective Potential}

The purpose of this section is to summarize briefly the main points
from Ref.~\cite{Mue02} which will, in turn, motivate our proposed 
method for constructing effective operators. To start, a bare 
one-dimensional $NN$ interaction with the same pathologies as a 
realistic interaction is assumed. That is, the bare potential is given 
by the sum of a strong short-range repulsive and a medium-range
attractive exponentials:
\begin{equation}
  V(x) =  V_{\rm s}\ e^{-m_{\rm s} |x|}
       +  V_{\rm v}\ e^{-m_{\rm v} |x|} \; .
 \label{VBare}
\end{equation}
The two masses were chosen to be equal to $m_{\rm s}\!=\!400$~MeV and
$m_{\rm v}\!=\!783$~MeV, respectively, while the strengths of the
potentials ($V_{\rm s}\!=\!-506$~MeV and $V_{\rm v}\!=\!+1142.49$~MeV)
were chosen to give a binding energy and point root-mean-square (rms)
radius for the symmetric (``deuteron'') state of 
$E_{\rm b}\!=\!-2.2245$~MeV and $r_{\rm rms}\!=\!1.875$~fm, 
respectively. 

Employing an option originally suggested by Lepage\cite{Lep97},
and later adapted by Steele and Furnstahl~\cite{Ste98,Ste99} to 
treat the $NN$ interaction, we propose a gaussian cutoff for the 
effective potential of the form
\begin{equation}
  V_{\rm eff}(x) = \frac{1}{a}\left(c
                 + d\frac{\partial^2}{\partial \xi^2}
                 + e\frac{\partial^4}{\partial \xi^4}
                 + \ldots\right)\exp(-\xi^2)\;;
                   \quad \xi\equiv x/a\;.
 \label{Veff}
\end{equation}
The parameters of the effective potential ($c$, $d$, $e, \ldots$) 
are fixed to reproduce the low-energy scattering phase shifts. That 
is, one adjusts the parameters until the following equation is satisfied:
\begin{eqnarray}
  k\tan\delta(k)
  \hspace{-0.05in}&=&\hspace{-0.04in}
  \frac{1}{a_{0}}-
  k^{2}\int_{0}^{\infty} dx\Big(\psi_{k}(x)\psi_{0}(x)-
                           \phi_{k}(x)\phi_{0}(x)\Big)\\
  \hspace{-0.05in}&=&\hspace{-0.04in}
  \frac{1}{a_{0}}-
  k^{2}\int_{0}^{\infty} dx\Big(\psi^{\rm eff}_{k}(x)
                                \psi^{\rm eff}_{0}(x)-
				\phi_{k}(x)\phi_{0}(x)\Big)\;,
\label{EffRange4}
\end{eqnarray}
where $\psi^{\rm eff}_{k}(x)$ is a scattering solution of 
Eq.~(\ref{SchrEq}) with $V(x)$ replaced with $V_{\rm eff}(x)$.
Note that as in Ref.~\cite{Mue02}, the gaussian cutoff parameter 
has been fixed at $a\!=\!1.16$~fm. The above condition may be 
rewritten in the following convenient form:
\begin{equation}
  \delta\langle{\mathcal I}\rangle(k;c,d,\ldots)\equiv
  \int_{0}^{\infty} dx\Big(\psi_{k}(x)\psi_{0}(x)-
       \psi^{\rm eff}_{k}(x)\psi^{\rm eff}_{0}(x)\Big)\equiv 0\;.
\label{Eff1}
\end{equation}
Evidently, it is neither demanded nor expected that Eq.~(\ref{Eff1}) 
be satisfied for arbitrary large values of $k$. Rather, one follows 
a hierarchical scheme, based on power counting, that assures that
observables in the bare and effective theory be indistinguishable at
low energies. It should be emphasized that the specific form of the
potential is somewhat arbitrary, as the short-range structure of the
theory becomes encoded in the effective parameters.

\begin{figure*}[ht]
\begin{center}
\includegraphics[width=5.0in,angle=0,clip=true]{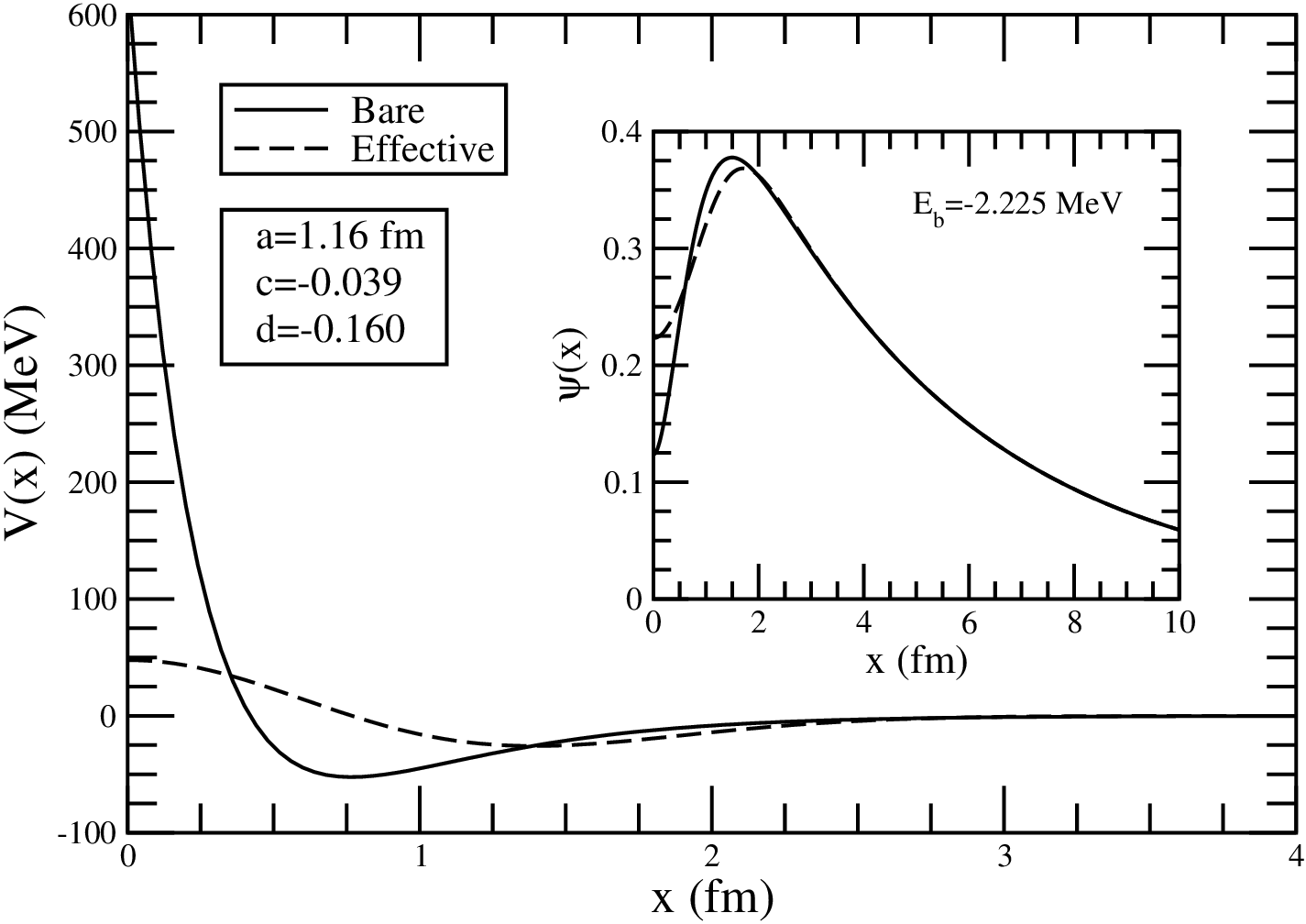}
\caption{Bare (solid line) and effective (dashed line) $NN$ 
potentials in a realistic 1-dimensional model. The inset shows 
``deuteron'' ground-state wavefunctions. The sharp features of 
the bare potential are no longer present in the effective 
potential. Although the short-distance structure of the 
wavefunctions are different, the exponential falloff (binding 
energy) is unchanged.}
\label{Fig1}
\end{center}
\end{figure*}

As a simple illustration of this procedure we display in
Fig.~\ref{Fig1} bare (solid line) and effective (dashed line) $NN$
potentials. The bare potential, with its characteristic short-range
structure, yields a scattering length of $a_{0}\!=\!5.247$~fm and an
effective range of $r_{0}\!=\!1.521$~fm, respectively. The calculation
of low-energy phase shifts is repeated using the effective potential
[Eq.~(\ref{Veff})] with its two parameters ($c$ and $d$) adjusted to
reproduce the exact effective-range expansion to order $k^{2}$. The
resulting parameters ($c\!=\!-0.039$ and $d\!=\!-0.160$) are natural
and yield a smooth effective potential which as far as the low-energy
properties of the theory are concerned, is practically indistinguishable
from the bare potential. Indeed, bulk properties of the ground-state
(henceforth referred to as the ``deuteron'') are predicted to be 
identical to those obtained in the bare theory. This in spite of the 
vastly different short-range structure of the wavefunctions 
(see inset in Fig.~\ref{Fig1}).

\subsection{Effective Operators}

One of the main criticisms levied on traditional shell-model
approaches is the lack of consistency between the construction of the
effective potential and the renormalization (if any) of the bare
operators~\cite{Hax99,Hax01}. While important steps have been taken to
correct this inconsistency, both in the area of low-dimensional
quantum magnets~\cite{Pie98b} and nuclear structure~\cite{Nav00b},
these are in the very early stages. Further, it is unknown how to
construct effective operators that are consistent with both the
effective theory and similarity-transformation based approaches.

In this contribution we propose an approach for the modification of
operators that is consistent, indeed mimics, the construction of
the effective potential. We assume that the momentum dependence of
matrix elements of (simple) single-particle operators may be accounted
for by an expansion in powers of $k^2$ having the same form as the
effective-range formula [Eq.~(\ref{EffRange3})]. To do so, one demands
that matrix elements of effective operators (${\cal O}_{\rm eff}$)
with scattering-wave solutions of the effective potential
($\psi_{k}^{\rm eff}$) possess the same momentum dependence as those
using the bare operators with the exact wavefunctions. In analogy 
with the definition of the effective potential [Eq.~(\ref{Veff})], we
parametrize the effective operators via
\begin{equation}
   {\mathcal O}_{\rm eff}(x;c,d,\ldots)=
   {\mathcal O}(x)
    \left[1+\left(c
                + d\frac{\partial^2}{\partial \xi^2}
                +  \ldots\right)\exp(-\xi^2)\right]\;.
 \label{EffOp}
\end{equation}
This parameterization affects only the short-range behavior of the
operator just as using the effective potential modifies only the
short-range structure of the wavefunction. The parameters $c$, $d$,
$\ldots$ (as before) are tuned to the low-$k^2$ behavior of the
exact matrix elements. To be more specific about our procedure, we 
fit---in complete analogy to Eq.~(\ref{Eff1})---the parameters of 
the effective operator by requiring that
\begin{equation}
  \delta\langle{\mathcal O}\rangle(k;c,d,\ldots)\equiv
  \int_{0}^{\infty}dx
     \Big(\psi_{k}(x){\mathcal O}(x)\psi_{k\!=\!0}(x)-
          \psi_{k}^{\rm eff}(x){\mathcal O}_{\rm eff}(x;c,d,\ldots)
          \psi^{\rm eff}_{k\!=\!0}(x)\Big)=0\;.
 \label{Eff2}
\end{equation}
The integral in this expression is convergent as the effective
theory demands that
\begin{subequations}
\begin{eqnarray}
  && \lim_{x\rightarrow\infty}\psi_{k}^{\rm eff}(x)=\psi_{k}(x)\;,\\
  && \lim_{x\rightarrow\infty}
 {\mathcal O}_{\rm eff}(x;c,d,\ldots)={\mathcal O}(x)\;.
 \label{fitb}
\end{eqnarray}
\end{subequations}
However, to insure that each term separately is convergent, we add 
and subtract the following term:
\begin{equation}
    \int_{0}^{\infty}dx
    \Big(\phi_{k}(x){\mathcal O}(x)\phi_{k\!=\!0}(x)\Big) \;,
 \label{fitc}
\end{equation}
where we recall that $\phi_{k}(x)$ is the free solution of the 1D
scattering problem [see Eq.~(\ref{BoundCondb})].

To extract the parameters of the effective operator we now 
fit---in the spirit of the effective-range expansion---the 
low-energy matrix elements of the bare operator between bare 
scattering wavefunctions according to 
\begin{equation}
  \langle{\mathcal O}\rangle_{\rm BB}(k)=
  \int_{0}^{\infty}dx
  \Big(\psi_{k}(x){\mathcal O}(x)\psi_{k\!=\!0}(x)-
       \phi_{k}(x){\mathcal O}(x)\phi_{k\!=\!0}(x)\Big)=
       \alpha+\beta k^{2}+\ldots
 \label{fitd}
\end{equation}
The parameters fixing the effective operators are then
adjusted so that the above expansion is recovered. That is,
\begin{equation}
  \langle{\mathcal O}\rangle_{\rm EE}(k)=
  \int_{0}^{\infty}dx
    \Big(\psi^{\rm eff}_{k}(x)
    {\mathcal O}_{\rm eff}(x;b,c,\dots)
         \psi^{\rm eff}_{k\!=\!0}(x)-
     \phi_{k}(x){\mathcal O}(x)\phi_{k\!=\!0}(x)\Big)=
    \alpha+\beta k^{2}+\ldots
 \label{fite}
\end{equation}
Note that when ${\mathcal O}(x)\!=\!{\mathcal O}_{\rm eff}\!=\!1$
one recovers the effective-range expansion.

\section{Results}
\label{results}

In this section we compute matrix elements of various operators using
three different schemes. The first scheme uses bare operators with
bare wavefunctions (we label these calculations as ``B+B''); these 
should be regarded as ``exact'' answers.  Second, we compute matrix 
elements in an approximation (labeled as ``B+E'') that uses effective
wavefunctions but retains bare operators. As we show below, for
operators insensitive to short-range physics this inconsistency
introduces small discrepancies. However, the more important the
short-range physics, the greater the lack of accord. Finally, we
perform calculations in a consistent low-energy approximation
(``E+E'') that employs both effective wavefunctions and effective
operators. Showing that these calculations are in excellent agreement
with the exact (B+B) answers represents the central result of the
present work.

Because of its simplicity, a natural place to start testing the 
proposed approach is the calculation of the root-mean-square radius 
of the deuteron, which is given by
\begin{equation}
 \langle{x}^{2}\rangle_{BB}\equiv
 \Big\langle\frac{1}{2}\sum_{n=1}^{2}
 \Big(x_{n}-x_{\rm cm}\Big)^{2}\Big\rangle=
 \int_{-\infty}^{\infty}dx
 \frac{x^{2}}{4}\psi_{\rm gs}^{2}(x)=
 (1.87977)^{2}~{\rm fm}^{2}\;.
 \label{XSquareBB}
\end{equation}
The corresponding calculation in the effective theory requires a
renormalization of the bare operator. To do so we follow the
prescription outlined in the preceding section [see Eqs.~(\ref{fitd})
and~(\ref{fite})] to obtain
\begin{equation}
  x^{2}_{\rm eff}=x^{2}
  \left[1+\left(c+d\frac{\partial^2}{\partial\xi^2}\right)
  \exp(-\xi^2)\right] \quad
  (c=1.520, d=-0.305)\;.
 \label{X2Eff}
\end{equation}
In this manner the root-mean-square radius predicted by the 
effective theory becomes
\begin{equation}
 \langle{x}^{2}\rangle_{\rm EE}=
 \int_{-\infty}^{\infty}dx\frac{x^{2}_{\rm eff}}{4}
 \Big(\psi_{\rm gs}^{\rm eff}(x)\Big)^{2}=
 (1.87988)^{2}~{\rm fm}^{2}\;.
 \label{XSquareEE}
\end{equation}
\begin{figure*}[ht]
\begin{center}
\includegraphics[width=5.0in,angle=0,clip=true]{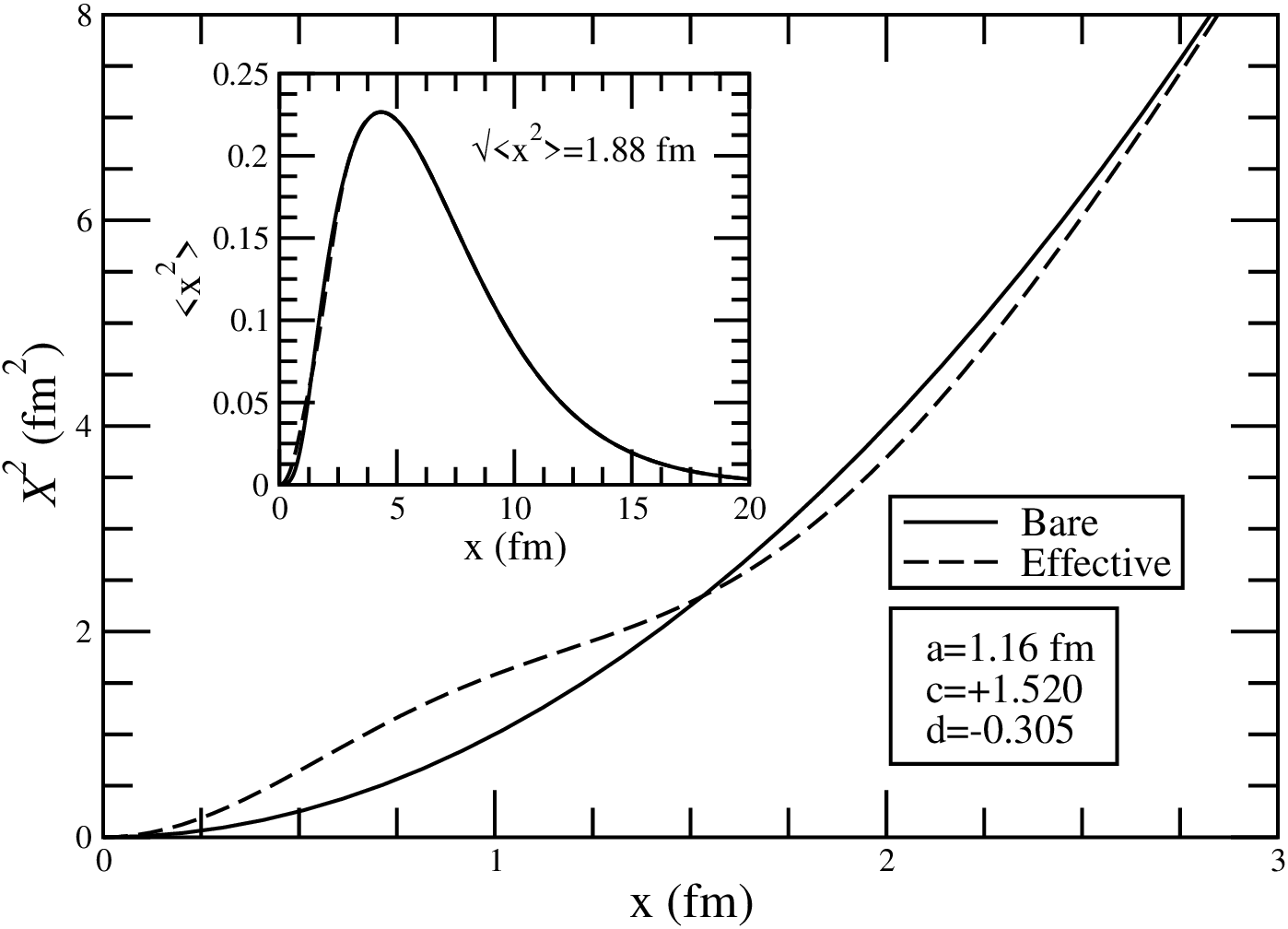}
\caption{Bare (solid line) and effective (dashed line) $X^2$ 
operator. Note that while the operators are considerably different 
at short distances, its ground-state expectation values are not
(see inset).}
\label{Fig2}
\end{center}
\end{figure*}
This represents a discrepancy of about $1$ part in $10^{4}$.  While
this result is gratifying and lends some credibiliity to the approach, 
it hardly qualifies as a stringent test of the formalism.  Although 
both the effective operator and the ground-state wavefunction are 
modified at short distances (see Fig.~\ref{Fig1} and~\ref{Fig2}) the 
operator itself has so little support at short distances that the 
two integrands [Eqs.~(\ref{XSquareBB}) and~(\ref{XSquareEE})] become
practically indistinguishable from each other (see inset on
Fig.~\ref{Fig2}). Indeed, an acceptable result is obtained even when
the operator is not properly renormalized:
\begin{equation}
 \langle{x}^{2}\rangle_{\rm BE}=
 \int_{-\infty}^{\infty}dx\frac{x^{2}}{4}
 \Big(\psi_{\rm gs}^{\rm eff}(x)\Big)^{2}=
 (1.87834)^{2}~{\rm fm}^{2}\;.
 \label{XSquareBE}
\end{equation}

\begin{figure*}[ht]
\begin{center}
\includegraphics[width=5.0in,angle=0,clip=true]{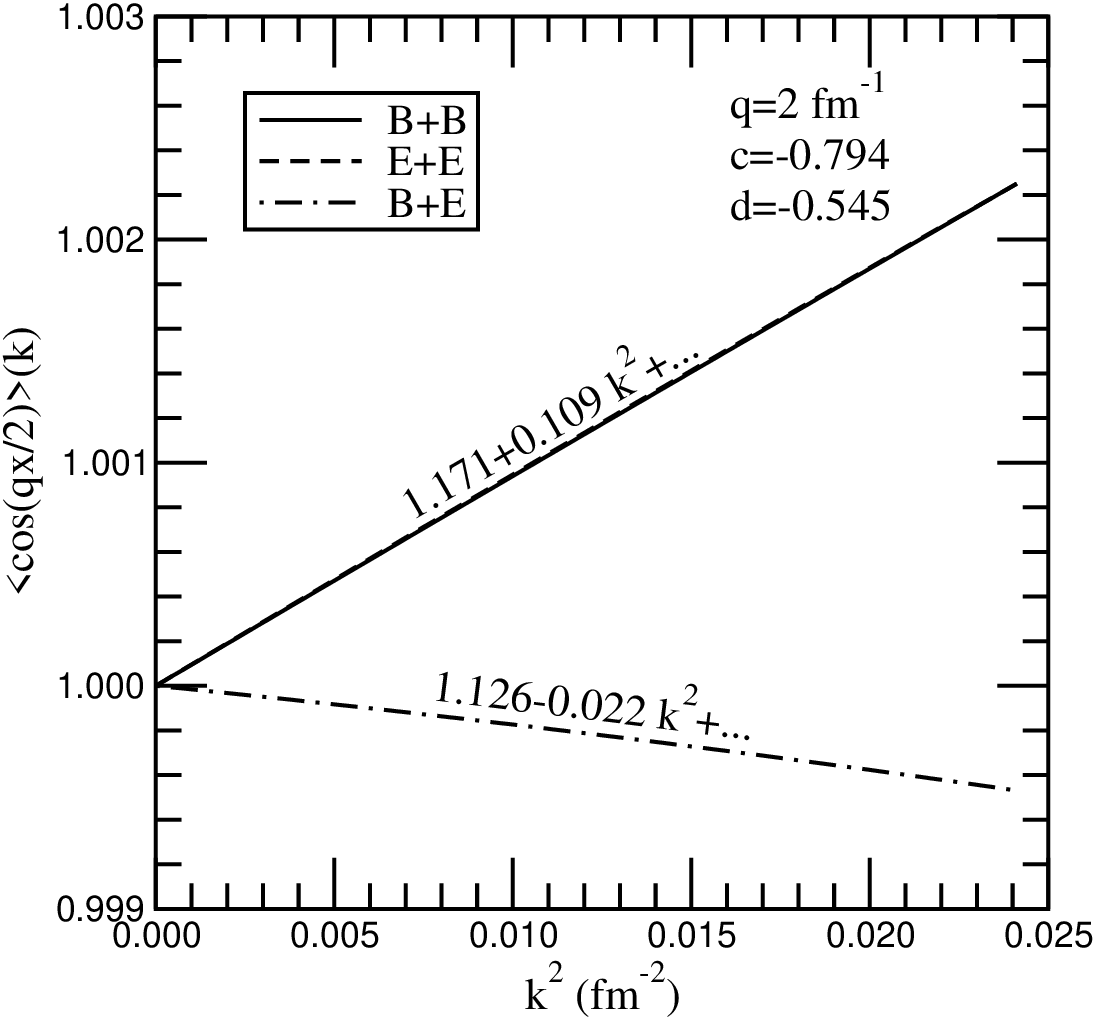}
\caption{An effective-range-like expansion for the elastic
form factor of the deuteron at $q\!=\!2~{\rm fm}^{-1}$.
Calculations are displayed for the bare theory (``B+B''),
the effective theory (``E+E''), and for a ``hybrid'' 
approximation that uses bare operators with effective 
wavefunctions (``B+E''). The solid (B+B) and dashed lines 
(E+E) are identical (by construction) since the effective 
parameters ($c$ and $d$) are tuned to reproduce the expansion 
for the bare theory.}
\label{Fig3}
\end{center}
\end{figure*}

\begin{figure*}[ht]
\begin{center}
\includegraphics[width=5.0in,angle=0,clip=true]{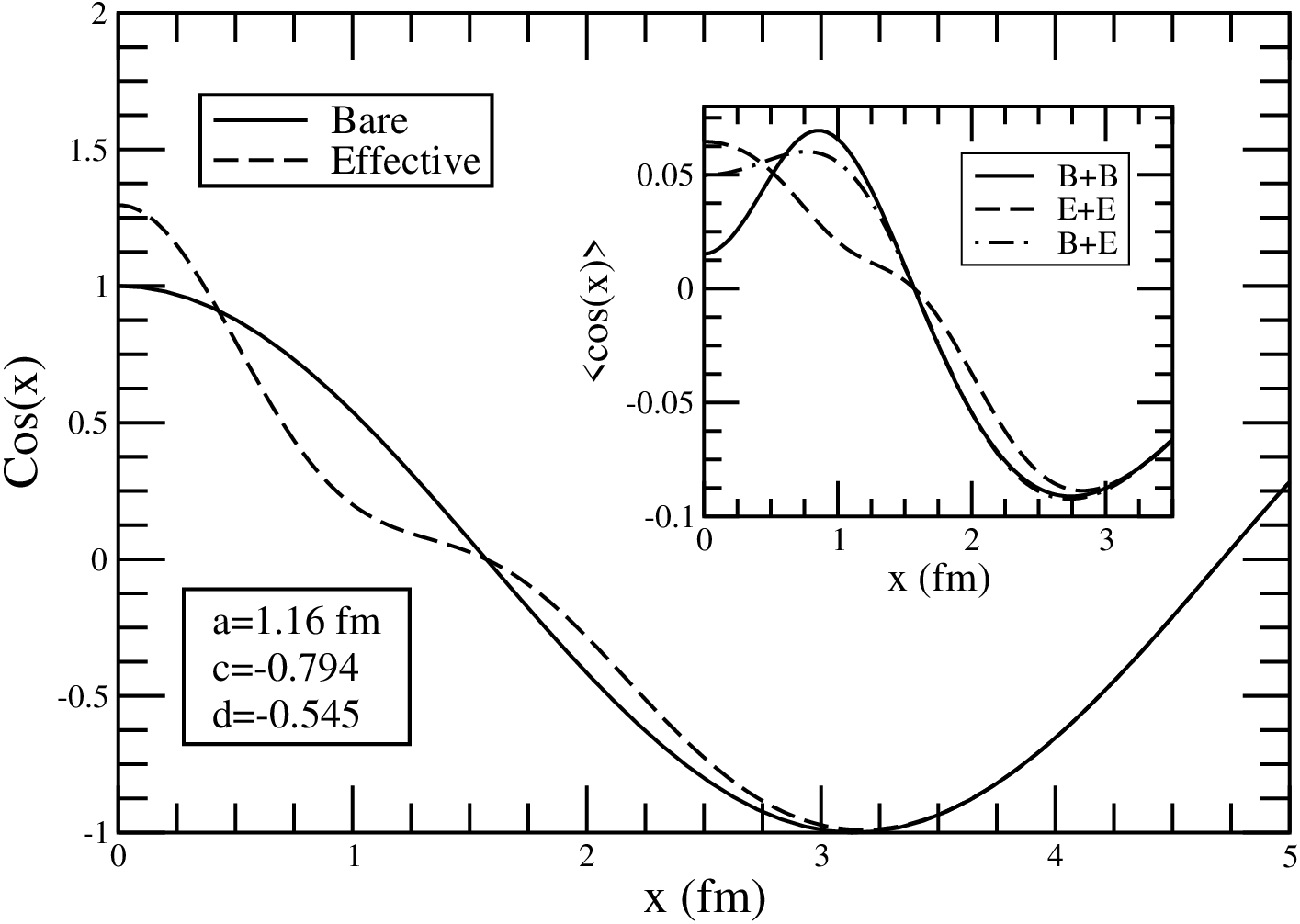}
\caption{Bare (solid line) and effective (dashed line) for
the $\cos(x)$ operator. The inset shows the product of the
square of the wavefunction with the operator for the three
calculations discussed in the text. The elastic form factor 
of the deuteron (at $q\!=\!2~{\rm fm}^{-1}$) is proportional 
to the square of the area under the curve.}
\label{Fig4}
\end{center}
\end{figure*}

\begin{figure*}[ht]
\begin{center}
\includegraphics[width=5.0in,angle=0,clip=true]{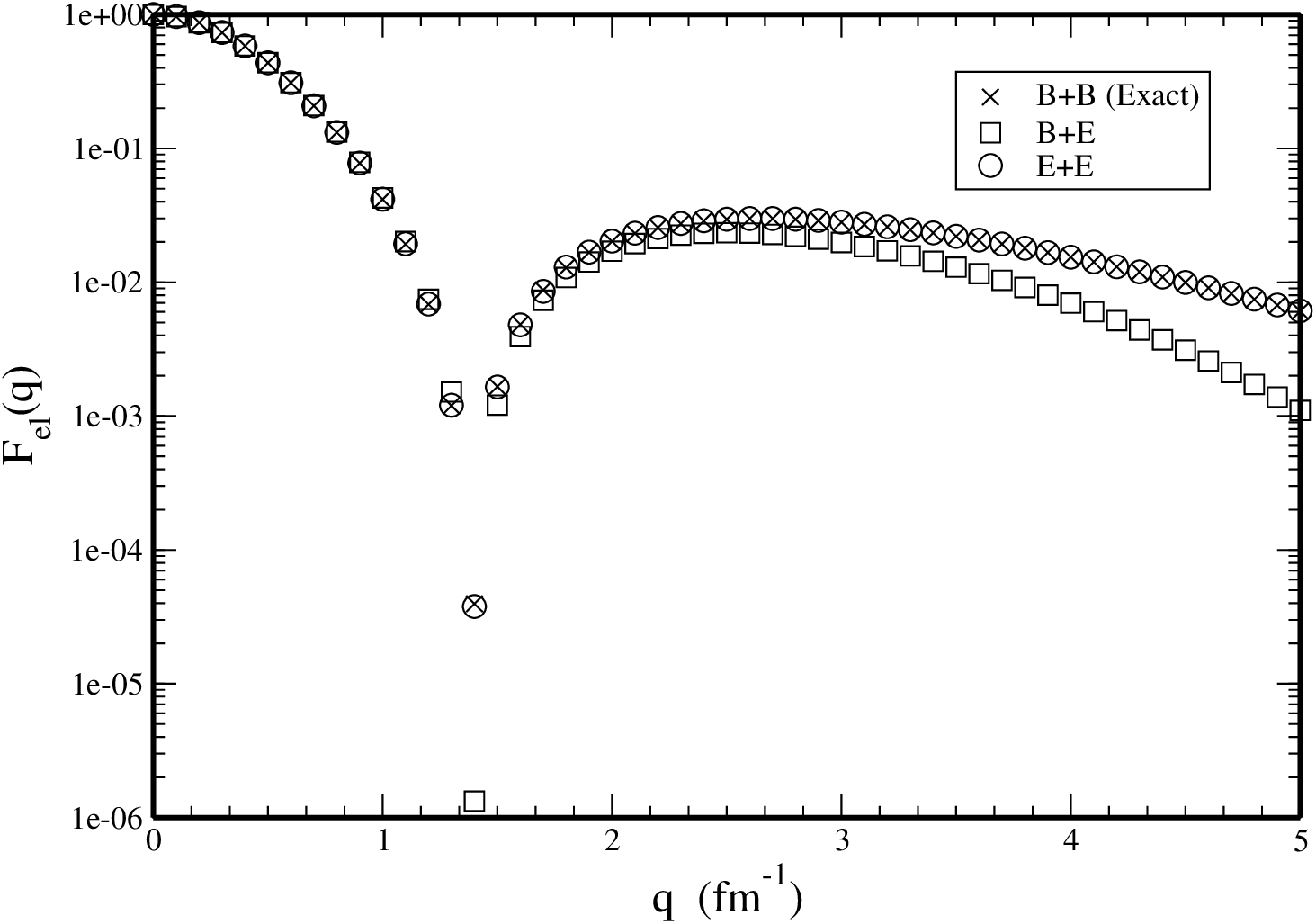}
\caption{The elastic form factor of the deuteron in the three
calculations discussed in the text. Note that the predictions
from the effective theory (E+E) agree with the exact theory at
at all values of the momentum transfer $q$.}
\label{Fig5}
\end{center}
\end{figure*}

A more sensitive test of the approach is provided by the elastic 
form factor of the deuteron, which in our simple one dimensional 
model reduces to the following expression:
\begin{equation}
 F_{\rm el}(q)=|\rho(q)|^{2}\;, \quad
 \rho(q)=\int_{-\infty}^{\infty}dx
 \cos\left(\frac{qx}{2}\right)\psi_{\rm gs}^{2}(x)=
 1-\frac{q^{2}}{2}\langle{x}^{2}\rangle+{\mathcal O}(q^{4})\;.
 \label{FormFactorBB}
\end{equation}
The corresponding expressions in the B+E and E+E approximations
are given by
\begin{subequations}
\begin{eqnarray}
  && \rho(q)_{\rm BE}=\int_{-\infty}^{\infty}dx
     \cos\left(\frac{qx}{2}\right)
     \Big(\psi_{\rm gs}^{\rm eff}(x)\Big)^{2} \;, 
 \label{RhoEB} \\
  && \rho(q)_{\rm EE}=\int_{-\infty}^{\infty}dx
     \left[\cos\left(\frac{qx}{2}\right)\right]_{\rm eff}
     \Big(\psi_{\rm gs}^{\rm eff}(x)\Big)^{2} \;,
 \label{RhoEE}
\end{eqnarray}
\end{subequations}
with the effective operator renormalized at short-distances as 
detailed above. That is,
\begin{equation}
  \left[\cos\left(\frac{qx}{2}\right)\right]_{\rm eff}=
  \cos\left(\frac{qx}{2}\right)
  \left[1+\left(c(q)+d(q)\frac{\partial^2}{\partial\xi^2}\right)
  \exp(-\xi^2)\right] \;,
 \label{CosqxEff}
\end{equation}
The renormalization procedure is illustrated in Fig.~\ref{Fig3}
at the single momentum-transfer value of $q\!=\!2~{\rm fm}^{-1}$.
It is important to note that the renormalization coefficients ($c$ 
and $d$) must be tuned at each value of the momentum transfer $q$. 
The solid line in the figure displays an effective-range-like 
expansion for the bare operator ${\mathcal O}(x)\!=\!\cos(qx/2)$ 
[as described in Eq.~(\ref{fitd}), with the slope and intercept 
clearly indicated in the figure (note that for clarity the plots 
are normalized to one at $k^{2}\!=\!0$). It becomes immediately 
evident that the predicted low-energy behavior of the exact theory 
can not be reproduced without a proper renormalization of the operator. 
Indeed, the B+E calculation predicts the wrong momentum dependence; 
the sign of the slope is wrong! In contrast, it becomes a simple
matter to tune the parameters of the effective operator to reproduce 
exactly the low-energy behavior of the exact theory (dashed line).
Having corrected the short-distance structure of the operator one
proceeds to compute the elastic form factor of the deuteron, which now
is a {\it prediction} of the effective theory. The structure of the
form factor (again at $q\!=\!2~{\rm fm}^{-1}$) is shown in
Fig.~\ref{Fig4}. The main panel shows bare ({\it i.e.,} $\cos(x)$) and
effective operators, displayed as solid and dashed lines,
respectively. Both the effective deuteron wavefunction (inset on
Fig.~\ref{Fig1}) and the effective operator differ considerably from
their bare counterparts at short distances---and so is the product of
the (square of the) wavefunction times the operator (inset on
Fig.~\ref{Fig4}). Yet the area under the curve---whose square is
proportional to the elastic form factor---is essentially unchanged.
For comparison, the exact (B+B) and effective (E+E) theories yield
values of $F_{\rm el}\!=\!0.02019$ and $F_{\rm el}\!=\!0.02012$,
respectively. In contrast, a (B+E) calculation with an effective
wavefunction---but still employing a bare operator---results in 
a discrepancy of nearly 20\% ($F_{\rm el}\!=\!0.017062$).
We conclude the discussion of the elastic form factor of the deuteron
by displaying in Fig.~\ref{Fig5} its momentum-transfer dependence up
to $q\!=\!5~{\rm fm}^{-1}$. Recall that effective parameters must be
tuned for every value of $q$. It is evident from the figure that the 
renormalization of the operator at high-momentum transfers is
essential, as it is at high $q$ that the short-distance structure of 
the wavefunction (and of the potential) is being probed. Failing to
correct the operator results in a rather poor representation of the
elastic form factor for $q\agt 2~{\rm fm}^{-1}$ (squares).

For completeness, and as a further stringent test of the formalism, 
we compute ground-state observables for an operator with the most 
extreme short-range structure possible: the Dirac delta function. 
In the bare theory the ground-state expectation value is simply 
given by the square of the deuteron wavefunction at the origin. 
That is,
\begin{equation}
 \langle\delta(x)\rangle=
 \int_{-\infty}^{\infty}dx\delta(x)\psi^{2}_{\rm gs}(x)=
 \psi^{2}_{\rm gs}(0)=0.01521 {\rm fm}^{-1}\;.
 \label{DeltaBare}
\end{equation}
As the sharp features of the bare potential are no longer present in 
the effective potential, the effective deuteron wavefunction at short
distances is considerably larger than the bare wavefunction (see 
Fig.~\ref{Fig1}). As a result, a (B+E) calculation using an effective 
wavefunction but a bare delta-function operator grossly overestimates 
the result: $\langle\delta(x)\rangle_{BE}\!=\!0.04981{\rm fm}^{-1}$.
Instead, by following the renormalization procedure outlined above,
one obtains an effective Dirac delta-function operator,
\begin{equation}
  \delta_{\rm eff}(x) = \frac{1}{a}
  \left(c+d\frac{\partial^2}{\partial\xi^2}\right)\exp(-\xi^2)
  \quad (c=0.250, d=-0.035)\;,
 \label{EffDelta}
\end{equation}
that yields a ground-state expectation value of
$\langle\delta(x)\rangle_{EE}\!=\!0.01523~{\rm fm}^{-1}$. This 
result deviates from the exact value by less than one part in a thousand.

In Table~\ref{Table1} we have listed (for completeness) some of the
results presented previously in graphical form. The operators appearing 
in this table are listed in order of the importance of their short-range
components. For example, the root-mean-square radius of the deuteron
depends little on the short-range structure of the wavefunction while
the Dirac $\delta$-function operator depends exclusively on it. For
each operator, we show the two effective coefficients $c$ and $d$
determined by the fitting procedure outlined above. All these are
dimensionless quantities and it is gratifying that they are all of
order one in keeping with the priciple of ``naturalness'' \cite{Kolck}.
We note
that in some cases failing to renormalize the operator (B+E) leads to
discrepancies that are as large as 50\% or even 100\%. In contrast,
calculations using effective wavefunctions and effective operators
(E+E) show excellent agreement with B+B calculations regardless of the
short-range structure of the operator. We stress that all these are 
predictions of the effective theory, as the tuning of parameters is 
done in the scattering sector. In particular, it is satisfying that 
the elastic form factor of the deuteron at $q\!=\!0$ deviates from 
unity by less than 2 parts in a thousand. We emphasize that such 
precise agreement is non-trivial; indeed, it reflects the soundness 
of our method.

\begin{table}[ht]
\centering
\medskip
\begin{tabular}{|c|c|c|c|c|c|}
$\langle{\mathcal O}(x)\rangle$ & $c$ & $d$ & B+B & B+E & E+E \\
\hline
$\sqrt{<x^2>}$ & $+1.51979$ & $-0.30545$ 
   & $1.87997$ & $1.87834$ (0.09\%) & $1.87988$ (0.01\%) \\
$|\langle\cos(0)\rangle|^{2}$ & $ -0.02016$ & $+0.02985$
   & $1.00000$ & $1.00000$ (0.00\%) & $1.00155$ (0.16\%) \\
$|\langle\cos(x/2)\rangle|^{2}$ & $-0.06440$ & $+0.13429$
   & $0.04159$ & $0.04261$ (2.45\%) & $0.04181$ (0.53\%) \\
$|\langle\cos(x)\rangle|^{2}$ & $-0.79386$ & $-0.54503$
   & $0.02019$ & $0.01706$ (18.4\%)&  $0.02012$ (0.35\%) \\
$|\langle\cos(2x)\rangle|^{2}$ & $-0.52925$ & $+0.14806$
   & $0.01541$ & $0.00698$ (220\%) & $0.01544$ (0.19\%) \\
$\langle\delta(x)\rangle$ & $+0.24964$ & $-0.03501$
   & $0.01521$ & $0.04981$ (327\%) & $0.01523$ (0.13\%)
\end{tabular}
\caption{Ground-state expectation values for various operators in 
the different approximation schemes described in the text. B+B 
indicates bare operators with bare wavefunctions, B+E bare operators 
with effective wavefunctions, and E+E effective operators with 
effective wavefunctions. Note that the root-mean-square radii are 
given in fm and $\langle\delta(x)\rangle$ in fm${}^{-1}$. Finally,
the $c$ and $d$ coefficients are dimensionless parameters of the 
effective theory.}
\label{Table1}
\end{table}

\section{Conclusions}
\label{conclusions}

While the field of nuclear structure has benefited from recent
advances in numerical algorithms and sheer computational power, the
shell-model problem, in its purest form, remains intractable.  As a
result, an important part of the nuclear-structure program for many
years has focused on the construction of effective interactions for
use in shell-model calculations. Two of the most promising approaches
are based on the so-called similarity transformation methods (in its
many varieties) and on effective-field-theory techniques.  The main
tenet underlying both approaches is that the short-distance structure 
of a theory (which is complicated and at present unknown) are hidden 
to a long-wavelength probe. It should then be possible to ``soften'' 
the corresponding short-range portion of the potential while leaving all
low-energy properties of the theory intact, thereby providing a 
significantly more tractible -- from a computational point of view --
interaction.

The main focus of the present paper is the determination of
single-particle operators which can be employed consistently in
conjunction with wavefunctions obtained using effective interactions.
As observed by many authors, such consistency is essential to the
correct implementation of effective theories. For computational
simplicity we adopted a one-dimensional $NN$ interaction, that
nevertheless incorporates the well-known pathologies of a realistic
$NN$ potential. The central result of this work is the proposal and
implementation of a single underlying approach for the construction of
both effective interactions and effective operators. The construction
of the effective interaction follows a well-known approach that is
based on a textbook derivation of the effective-range expansion.
What is not well known (at least to us) is that the same approach
may be generalized to effective operators.

Results from such an implementation are very gratifying, as evinced
from a variety of calculations of ground-state observables. For those 
observables insensitive to the short-range structure of the potential, 
such as the root-mean-square radius of the deuteron, the
renormalization of the bare operator, while required by consistency,
is of little numerical consequence. Yet, failing to properly
renormalized operators sensitive to short-range physics, such as the 
elastic form factor of the deuteron at high-momentum transfers, 
can yield discrepancies as large as 200\%. The consistent
renormalization procedure advocated here yields in all cases errors 
of less than 1\%. 

We conclude with a short comment on future work. The results presented
here are encouraging and lend validity to the proposed approach, which
is currently being extended to the three-body system. Different
algorithms are being employed to solve for the ground-state of the
three-body system and in all cases, perhaps not surprisingly, better
convergence properties are obtained with the effective rather than
with the bare interaction. The results obtained here also constitute a
promising first step toward our ultimate goal of combining
similarity-transformation methods with effective interactions. The
effective interactions and operators obtained here---with their sharp
short-range features no longer present---could provide a more suitable
starting point for the numerically-intensive approaches based on
similarity transformations. Finally, the method proposed here will
have to be extended to three-spatial dimensions. Other than numerical
complexity, we do not foresee other serious challenges. Indeed, the
approach presented here for the construction of effective interactions
(whose three-dimensional derivation may be found in several textbooks)
had to be adapted to one spatial dimension. In summary, a novel
approach for the renormalization of operators in a manner consistent 
with the construction of the effective potential has been proposed
and implemented with considerable success. The results obtained here
are gratifying and suggest how in the future effective theories may be
profitably combined with more traditional methods to tackle the 
nuclear many-body problem.

\acknowledgments
This work was supported in part by the U.S. Department of Energy
under contracts DE-FG05-92ER40750 and DE-FG03-93ER40774.

 
\end{document}